# Emergence of the Traffic Autonomous Zone (TAZ) for Telecommunication Operations from Spatial Heterogeneity in Cellular Networks

Liyan Xu[1, *], Jintong Tang[2], Hezhishi Jiang[3], Hongbin Yu[1], Yihang Li[4], Qian Huang[4], Yinsheng Zhou[4], Jun Zhang[5], Yu Liu[1]

1. College of Architecture and Landscape Architecture, Peking University, Beijing, China.
2. Institute of Remote Sensing and Geographical Information Systems, School of Earth and Space Sciences, Peking University, Beijing, China.
3. Academy for Advanced Interdisciplinary Studies, Peking University, Beijing, China.
4. Global Technical Service Dept, Huawei Technologies, Beijing, China.
5. Global Technical Service Dept, Huawei Technologies, Shanghai, China.
[*]: Corresponding Author.

# Abstract

In the field of telecommunications, various operations are driven by different physical quantities. Each has its own patterns in time and space, but all show some clustered structures in their spatial distribution. This reflects a unified rule of human mobility, suggesting the consistency among different telecommunication regionalization objectives. With this in mind, regionalization can be used to identify these patterns and can be applied to improve management efficiency in the context of "autonomous networks". This article introduces the "Traffic Autonomous Zone (TAZ)" concept. This approach aims to create a reasonable unified regionalization scheme by identifying spatial clusters. It is not just a practical way to partition cities based on telecommunications needs, but it also captures self-organization structure of cities in essence. We present examples of this regionalization method using real data. Compared to the popular Louvain community detection method, our approach is on the Pareto frontier, allowing for a balance among various metrics in telecommunications.

# 1 Introduction

Autonomous Networks (AN) [1], or Autonomous Driving Network (ADN) [2] are an important concept for next-generation cellular networks, which is defined as "a telecommunication system (including management system and network) with autonomy capabilities which is able to be governed by itself, with minimal to no human intervention" [3]. Given the extremely large scale of cellular networks, achieving the overall network autonomy is operationally challenging, especially for high-level ANs. However, the logical and spatial hierarchical structure of cellular networks suggests a possible "divide-and-rule" path for AN construction. Indeed, cellular networks serve as a whole, while each base station covers only a limited area around it; further, a group of neighboring base stations serves a specific geographic area and shares infrastructure such as transmission line, spare parts depots, and maintenance personnel. These features map the hierarchical structure of the cellular network to certain clustered structure in space.

The demand for a "divided-and-ruled" network necessitates a reasonable geographical regionalization of the service areas, an analysis with a long-standing academic tradition in geography. Conceptually, the goal of geographical regionalization is to divide continuous space into several reasonably shaped and scaled regions. Each region should have consistent characteristics internally, while distinct differences should exist among regions [4]. This facilitates the implementation of targeted measures based on the unique characteristics of each region. Geographical regionalization is widely applied, with common examples including administrative divisions [5], electoral districts [6], and urban planning and management zones [7]. In fact, regionalization already has applications within the telecommunications field. Some operators divide cities into different "micro-grids" based on service scenarios, providing differentiated network planning and facilities to achieve specific management effects [8], [9].

Despite these advancements, the delineation of geographical regionalization for telecommunication services still largely relies on ad hoc experiences. This brings about two main issues. On the one hand, variations in individual experiences result in a lack of unified technical standards for regionalization, making it difficult to replicate on a larger scope. On the other hand, personal experiences are not always reliable and do not guarantee accurate results. Establishing a unified methodological framework is thus crucial, yet challenging. A unified telecommunication regionalization involves multiple criteria covering data traffic, human mobility, and place semantics. Each of the objectives further involves particular performance indicators with the corresponding physical quantities differ in spatial distribution terms, and thus optimal regionalization solutions for different objectives may conflict. Note that the regionalization is also constrained by urban structure, morphological features, and various operational considerations, making it a multi-objective, multi-constraint problem. As a result, a unified regionalization approach inevitably involves trade-offs among multiple goals. Moreover, it is inevitably affected by the modifiable areal unit problem (MAUP) [10]. It stems from the continuous nature of geographical variables and the instability of their heterogeneity, and may render any geographical regionalization "wrong" [11], or rather, not the optimal solution to the problem. Facing such a "wicked problem" [12], one has to in the first place ensure a suboptimal solution that at least avoid blatant errors like gerrymandering [13] is pursuable, which would require certain degree of consistency among the relevant characteristic scales and forms of geographical variables involved.

Fortunately, as a behavioral mapping of human mobility across space, the spatial distribution of major telecommunication operation and service indicators should conceptually correlate highly with the statistical physical laws of the former. These laws have been extensively understood, including the nested spatial container structure of human activities [14], the regularity and predictability of human movement patterns [15]–[18], the universal regularities of urban scale and morphology [19], and the scaling laws of human behavior in urban complex systems [20]. What is worth noting in these findings is that the spatial self-organizing characteristic scales of human activities are coupled with those of cellular networks. Indeed, the service radiuses of base stations vary from about 2 kilometers (for 2G networks) to 200 meters (for 5G networks) [21], [22], and those of base station clusters are determined accordingly. Meanwhile, by geographical laws [23], human activities, including aggregation, movement, and other semantic behavior, exhibit heterogeneity at comparable spatial scales [11], [24], forming an archipelago-like spatial structure where each "island" has distinct behavioral signature, and this in turn is "translated" to unique telecommunication service demands by the direct mapping relationship between human behavior and telecommunication activities [25]. These include hot and cold spots in the spatial distribution of network traffic [26], heterogeneous demand zones for cellular service experiences arising from semantic features of places such as urban functions [27], etc., all corresponding to certain Key Performance Indicators (KPIs) and Key Quality Indicators (KQIs) in network planning and optimization such as coverage, capacity, and stability, which serve as proper regionalization indicators.

Overall, cellular networks possess certain spatially localized self-organizing structure not only in terms of operations but also in terms of services, which not only highlights the need for regionalized operations of the entire cellular network, but also inspires us to hypothesize that there is potentially an innate consistency among the multiple objectives of telecommunication regionalization, such that a reasonable unified regionalization scheme allowing local autonomy of the network at certain spatial scales is feasible. Such a regionalization would reflect the self-organization characteristics of key telecommunication operation and service indicators in space, and hence is capable of serving users with common features and needs through a unified solution within a certain spatial scope, which would help the establishment of ANs.

Given the foundational role of cellular traffic in contemporary telecommunication services, we term this regionalization scheme as the "Traffic Autonomous Zone (TAZ)". It is defined as a telecommunication operational and service regionalization scheme that meets the following criteria:

1. It should reflect the inherent self-organizing spatial structure of key performance indicators involved in telecommunications such, rather than being an arbitrary designation. Particularly, it must consider a series of KPIs and KQIs in telecommunication operations, including user capacity within the region, interactions among regions, urban place semantics of the regions, and the morphology of the natural and built environment.
2. It should encompass the full process of various telecommunication operations, such as network planning, construction, maintenance, optimization, and marketing. This means that, in addition to the aforementioned KPIs and KQIs, the main indicators of the above services should also be included in the regionalization objectives.
3. It should have consistent regionalization logic and determinate regionalization results, while allowing for necessary operational flexibility. This requires that the TAZ, while considering the needs of the various types of operations mentioned above, allows for prioritization and trade-offs among multiple potential objectives. However, the final spatial regionalization approach should be unique to satisfy the convenience requirements of day-to-day practices.

Evidently, these criteria dictate that the TAZ is both discovered (Criterion 1) and constructed (Criteria 2 and 3), thus balancing the reliability of scientific foundations with practical demands. In the remainder of this paper, we will introduce the theoretical foundation and delineation methods of the TAZ, accompanied by an application example. Section 2 is dedicated to the analysis of the spatial distribution laws of the main physical quantities at stake and the interrelationships among them, thus to provide a conceptual basis for the theoretical existence of the TAZ. Section 3 will present the formulation and solutions for the regionalization problem. Section 4 will showcase the delineation results of the TAZ through an example and evaluate its performance. Finally, Section 5 summarizes the entire paper.

# 2 Performance Indicators in Telecommunications and Related Physical Quantities: Feasibility of A Spatial Compromise

This section will enumerate the relevant physical quantities that correspond to major performance indicators in telecommunications, and analyze the fundamental spatial-temporal laws they follow that have regionalization implications, as well as the interrelationships between different physical quantities that allow for a spatial compromise. These analyses will demonstrate that the delineation of the TAZ is not an arbitrary construction, but rather reflects the essence of the spatial-temporal laws in telecommunication.

## 2.1 Physical Quantities That Correspond to Telecommunication Performance Indicators

Mobile communication service and operations involve multiple practices including the planning, construction, maintenance, optimization, and marketing, each with distinct KPIs (for operators) and KQIs (for customers) that correspond to specific physical quantities [28], [29]. We summarize these performance indicators and the related physical quantities, as well as the resulting regionalization objective by service/operation scenarios in Table 1.

Table 1   Telecommunication performance indicators and the corresponding physical quantities

| Service/ operation scenario | Service/ operation activity | Spatiotemporal characteristics | Physical quantity | KPI | KQI | Regionalization objective |
|---|---|---|---|---|---|---|
| User communication activity at rest | Upstream and downstream transmission between terminal and base station | Low-order spatiotemporal features: spatiotemporal distribution of | Population, call volume, traffic, etc. | Transmission byte, access drop rate, successful establishment | Service availability, response time, connect time, video/audio | Stable area for "static", low-frequency spatiotemporal patterns of the |

| | | physical quantities | | rate, successful connect rate, etc. [30] | quality, speech quality, end-to-end delay, etc. [29] | population |
|---|---|---|---|---|---|---|
| Communication activities while the user is on the move (e.g., while traveling in a car) | Handovers between base stations or cells during movement | High-order spatiotemporal features: mobility and interactions | User movement trajectory | Handover successful rate [31] | Handover latency [32] | Stable community of spatial interaction network |
| High-value scene identification, precision policymaking according to area themes, marketing to specific customers | Various types of regular communication activities for specific customers or places | Specific combination of low-order and high-order spatiotemporal features: typical communication usage patterns and service demands | Place semantics | Depending on the particular service needs of the place | Depending on the particular service needs of the place | Clearly defined thematic areas for urban functions |
| Operations requiring personnel on-site | Survey, construction, troubleshooting, etc. | Low-order spatiotemporal features: spatial patterns and boundaries | Urban morphology, natural features, administrative boundaries, etc. | Managing costs [33] | Mean time to repair (MTTR), etc. [34] | Clearly defined management zones |

### 2.1.1 Service Coverage, Capacity, and First-order Spatiotemporal Quantities: Population and Cellular Traffic Distribution

Capacity planning is an important step in network planning. It requires estimating the local traffic profile, which includes the scale of users, types of services, traffic usage, and so on. Subsequently, it is necessary to calculate metrics like throughput per user to forecast the local planning capacity and equipment scale [35]. For the telecommunication regionalization task, its possibility relies on two crucial conditions, namely, (1) the existence of local population clusters which follows general statistical physical laws; and (2) the spatial correspondence between cellular traffic and population distribution.

For the former, geographical research has long focused on the spatial distribution of urban population. Early classical studies revealed population distribution models for monocentric cities, such as the negative exponential form presented by the Clark model [36] and quadratic exponential forms [37]. Later research quantitatively demonstrated that a polycentric model better characterizes urban population distributions [38] and introduced mathematical models for the population distribution in cities with polycentric structures [39].

For the latter, studies on mobile phone signaling discovered that communication activities within cities exhibit either monocentric or polycentric distribution characteristics, corroborating the findings for urban population. Moreover, larger cities tend to have more hotspots, and their spatial distributions are more inclined to show a polycentric structure, and this finding is robust across changes in spatial grid aggregation scales [40]. Similarly, the spatial distribution of traffic loads within cities exhibits clear heterogeneity. Studies have revealed that the traffic load and its density in a few areas are significantly higher than in most areas [41]–[43]. This seems to reflect a ubiquitous scaling law in urban science [44]. Furthermore, some research has proven that, based on mobile phone data and appropriate conversion methods, one can effectively estimate the distribution of population density [45]. This confirms the aforementioned mapping relationship between human activities (spatial aggregation) and telecommunication metrics.

In summary, the above research indicates that the spatial distribution laws for various physical quantities like population and cellular traffic within cities are consistent. This establishes the possibility of creating reasonable regional divisions based on the distribution patterns of these indicators, or "first-order" spatiotemporal quantities [46].

### 2.1.2 Dynamic Communication Stability and High-order Spatiotemporal Quantities: Human Mobility and Spatial Interaction

In mobile communications, there is an inherent need for "mobility", which aims to minimize unnecessary handovers and handover failures during a user's movement, ensuring stable communication [47]. This requirement is particularly crucial for 5G networks, as 5G base stations have smaller cell coverage areas and thus handovers are more frequent. Spatially, a natural solution is to delineate a service zone where a fixed group of base stations ideally covers most of the spatial movement range of the majority of users, thereby reducing the frequency of handovers. This allows for holistic optimization of base station parameters within these zones to accommodate mobility needs.

The underlying physical context of the above process is the laws of human mobility and spatial interaction. Unlike the "first-order", or static distribution discussed in the previous section, human mobility represents a "higher-order", or dynamic process [46]. It is more complex, but it nevertheless also follows certain general laws. Based on the behavioral foundation of human mobility [15], [18], [48], researchers have proven the high predictability (at least 93%) of human mobility patterns [16]. Given a distribution of movement Radius of Gyration (RoG) for a group of people, one can estimate the spatial activity range for a large number of users [17]. Furthermore, the range of human movement displays a hierarchical structure in space [14]. These two findings not only imply that given a population, one can identify the appropriate scale covering its activity area, but they also indicate the existence of a very limited set of such characteristic scales.

In the context of telecommunications, this means that multilevel regionalization based on human mobility is feasible, where each spatial level's regional unit covers a unique and much fixed user group, hence allowing for targeted fine-tuning of the network. Technically, we can identify such regions by performing community detection on the spatial interaction network formed by human mobility. What is fortunate is that studies have shown that community structures discovered in such spatial networks often display geographical contiguity [49], [50], which could be a direct consequence of the distance decay principle of the intensity of spatial interactions [23], [51]. Moreover, communities identified through spatial interactions tend to be multi-layered and nested, which usually correspond to the hierarchical structure of the city per se [52], another good news for operational convenience which we will cover later.

### 2.1.3 Service Experience Quality and Place Semantics

In urban spaces, there exist certain "places" which are defined as "spatial locations that have been given meaning by human experiences" [53] in geography. Or more technically, place semantics refer to the functional use of urban land, such as transportation, work, dwelling, and recreation [54], or further detailed functions. Place semantics directly determine the behavior of people within them, leading to variations in their telecommunication service needs, especially in terms of experiential demands. For instance, train stations, with their dense crowds and frequent comings and goings, require maintaining high load capacity and frequent user access operations. In network planning, different communication activity characteristics necessitate the adoption of various service models, establishing a connection between place semantics and service models. For example, 5G network service models are categorized into three types: enhanced mobile broadband (eMBB), massive machine-type communications (mMTCs), and ultra-reliable low-latency communications (URLLCs). Within these, there are numerous subdivided application scenarios, each with distinct communication characteristics and capability requirements, correlating with different place semantics [55]. Therefore, proper identification of place semantics is crucial for telecommunication services.

The concept of "place" is ambiguous and subjective [56]. Therefore, the identification of places often lacks a standard answer. Despite this, at the operational level, past research has provided inspiring methods for identifying place boundaries and their semantics. Firstly, although "semantics" are ambiguous, certain places expressing specific functions still have boundaries due to the "top-down" delineation of urban planning. Intuitive examples include parks, schools, etc., which are established on specific plots of land and therefore have clear spatial extents. Secondly, even for areas with ambiguous boundaries or places with mixed functions such as commercial districts, one can still identify the commonly recognized semantics of "places" (possibly having multiple semantic labels simultaneously) and their boundaries using various Social Sensing data sources [57]–[59] by "bottom-up" methods. Lastly, the "semantics" of a place corresponds to a limited scale. Current research identifies "semantics" at scales below the city level, such as neighborhoods, plots, buildings, etc. These characteristics make it feasible to delineate regions within a city based on place semantics.

### 2.1.4 Operability of Regionalization and Urban Morphology

In practical terms, as the direct goal of delineating the telecommunication management zones is to guide on-site operations, including surveying, construction, maintenance, and operation, such regionalization should ideally be conveniently operable. Particular requirements include: (1) the size of the zone should neither be too large to ensure the workload of the on-site personnel at a reasonable level, nor too small for obvious reasons; (2) the zones should be morphologically regular enough such that the on-site personnel can cover the zone in a short time without spending excessive time on the road; (3) the boundaries of the zones should be intuitively understandable with reasonable common sense and domain knowledge; and (4) the zones should not pose significant challenges for management operations, such as coordinating with multiple administrative jurisdictions within one zone, or having internal accessibility barriers due to geographical obstructions like major rivers. To achieve these goals, natural geographical features, urban morphology, and administrative divisions must be taken into account as constraints in telecommunication regionalization.

Fortunately, in real-world cities, these constraints are generally not arbitrary. Taking urban morphology as an example, there is a bidirectional relationship between it and human activity. It is found that the street-block morphology in cities worldwide exhibit a high degree of regularity with only four prototypes [60]. Urban morphology also influences human activity. For example, within a city, the circulation of people, goods, and information is generally obstructed by physical features like major roads, railways, mountains, and rivers. These features create boundary effects on human activity [61]. It can be said that a regionalization scheme reflecting urban morphological characteristics is essentially capturing the spatial features of human activity within the city. Therefore, at least conceptually, these constraints are inherently consistent with the aforementioned regionalization requirements.

## 2.2 Compatibility and Trade-offs among the Physical Quantities, and Feasibility of Unified Regionalization

The four dimensions of indicators, i.e., quantity, interaction, semantics, and morphology, exhibit inherent geographic consistency or compatibility. This can be argued from two perspectives for any pair of physical quantities: (1) There exists a certain mapping mechanism between the two physical quantities. (2) When combined, the two physical quantities can form a spatial structure through either trade-offs or coupling, which manifests as balanced partitioning in regionalization tasks.

(1) Quantity-Interaction: For the mapping relationship between the two, studies deriving interactions from quantities are represented by the classic gravity model [62], including the intervening opportunities model [63] and the Exploration and Preferential Return (EPR) model [15]. These mechanisms have led to specific inference methods, such as deriving population flows from continuous quantity snapshots [64]. Mapping interaction to quantity typically requires initial values. If interaction and quantity represent different states of the same entity (e.g., population staying vs. moving), quantity values can be directly calculated from initial quantity values and movement quantities. If not, predictive models like STGCN and its derivatives use graph convolutional networks to predict future quantity values by constructing interactions as graph structures [65]. For balanced regionalization derived from both, examples include balancing commercial site visits within regions and flow intensity between regions [66], and balancing urban patrol zones based on historical cases and travel times [67].

(2) Semantics-Interaction: Functional differentiation of spatial units is the root cause of different types of interactions [46]. The Deep Gravity model can capture nonlinear relationships between semantic information (e.g., land use) and interactions [68]. Conversely, examining interactions over time can infer place semantics [69], [70], demonstrating a bidirectional mapping relationship. For balanced regionalization combining the two, the effectiveness varies with interaction intensity and semantic randomness. Real-world urban networks, a mix of strong and weak spatial constraints [71], and semantic distributions, a mix of highly functional and highly mixed areas [72], [73], provide conditions for balanced partitioning[74].

(3) Semantics-Morphology: Quantifiable relationships exist between urban block morphology and land use types [75]. Morphology can infer semantic differences, such as using geographic features (e.g., rivers, railways) to deduce socioeconomic segregation [76] or using building morphology to infer semantic differences [59], [77]. For balanced regionalization, Chodrow proposed a regionalization method balancing ethnic differences and morphological regularity, demonstrating the feasibility of semantic-morphology balanced partitioning [78].

(4) Quantity-Semantics: Phenomena show that quantity values and temporal curves can infer semantics [59], [79]. Mechanistically, bid-rent curves can explain the spatial clustering of land use types in monocentric cities [80], while the Fujita-Ogawa model [81] and the edge city model [82] can explain polycentric urban distributions, depicting the mapping mechanism between quantity and semantic distributions.

(5) Quantity-Morphology: The Max-p-compact regions problem addresses the need to balance quantitative attributes and partition morphology [83].

While not exhaustive due to space constraints, the listed combinations clearly exhibit transitive relationships.

Despite the potential geographic compatibility, electromagnetic laws rule that certain objectives in telecommunication optimization are intrinsically mutually exclusive. For example, Y. Chen et al. (2011) proposed four sets of trade-offs in wireless communication: deployment efficiency-energy efficiency trade-off, spectrum efficiency-energy efficiency trade-off, bandwidth-power trade-off, and delay-power trade-off. Such trade-offs are commonplace in network planning and optimization. For instance, during the scheduling process of 4G LTE network, limited wireless resources are allocated to users through a scheduling process. For the entire system, the goal is to maximize the system's total throughput; for individual users, fairness among users is sought; and for a single user, the aim is to meet their quality of experience. Existing scheduling algorithms strike a balance among these objectives [85].

On balance, a common behavioral foundation establishes an inherent possibility for unified regionalization in telecommunication operations. However, the differentiated objectives of different service categories introduce a need for trade-offs. Therefore, a multi-objective optimization framework is essential in spatial regionalization of telecommunications.

# 3 Problem Statement and Solution

## 3.1 Problem Statement: A Multi-objective Optimization Framework

Based on the analysis in Section 2, the TAZ partitioning problem is formally stated as the following multi-objective optimization problem:

$$max_Z\{f_i(Z), i = 1, ..., m\} \quad (1)$$
$$s.t. \quad Z = z(\Theta) = \{Z_C^j, j = 1, ..., n\}, \quad (2)$$
$$\Theta = \{param_l, l = 1, ..., p\}, \quad (3)$$
$$Z_C^j = \{Z_U^k, k = 1, ..., k_j\}, \quad (4)$$

$$Z_U = \{Z_U^r, r = 1 \ldots, q\}, \tag{5}$$

$$A_{min} \leq A(Z_U^r) \leq A_{max} \tag{6}$$

where equation (1) is the optimization objective. $Z$ refers to the solved TAZ scheme, which is the decision variable of the problem. $f_i(Z)$ is the $i^{th}$ optimization objective for $Z$. There are $m$ objective functions in total.

Constraint (2) states that $Z$ can be regarded as a set containing $n$ zones mathematically, each of which is denoted as $Z_C^j$. In terms of functions, $Z$ is actually the dependent variable of a partition function with the set of regionalization parameters $\Theta$ as the independent variable of the regionalization function $z: \Theta \to Z$. Constraint (3) gives the composition of the parameter set $\Theta$, where $param_l$ is the $l^{th}$ regionalization parameter, and there are $p$ regionalization parameters in total. Constraint (4) means that each region is a set of partition units $Z_U$, and the region $Z_C^j$ contains $k_j$ regionalization units. The $k^{th}$ regionalization unit in the partition is denoted as $Z_U^k$. Constraint (5) gives $Z_U$, the set of all regionalization units for the entire study area, where $Z_U^r$ is the $r^{th}$ regionalization unit of the entire study area, and there are $q$ regionalization units. Constraint (6) is the scaling constraint, where $A(Z_U^r)$ is the area of the regionalization unit $r$. $A_{min}$ and $A_{max}$ are the minimum and maximum areas of the participating regionalization units. The objective of this multi-objective optimization problem is to find the TAZ regionalization scheme that achieves Pareto optimality on the objective function.

We designed a regionalization methodology as shown in Fig. 1. Raw data are converted to the regionalization unit level to form the characteristics of each regionalization unit, and the regionalization scheme is formed through multilevel network community detection, and kernel extension [74]. Based on the regionalization units and regionalization scheme, we designed several multilevel regionalization technology routes. In the following section we briefly describe the entire methodology.

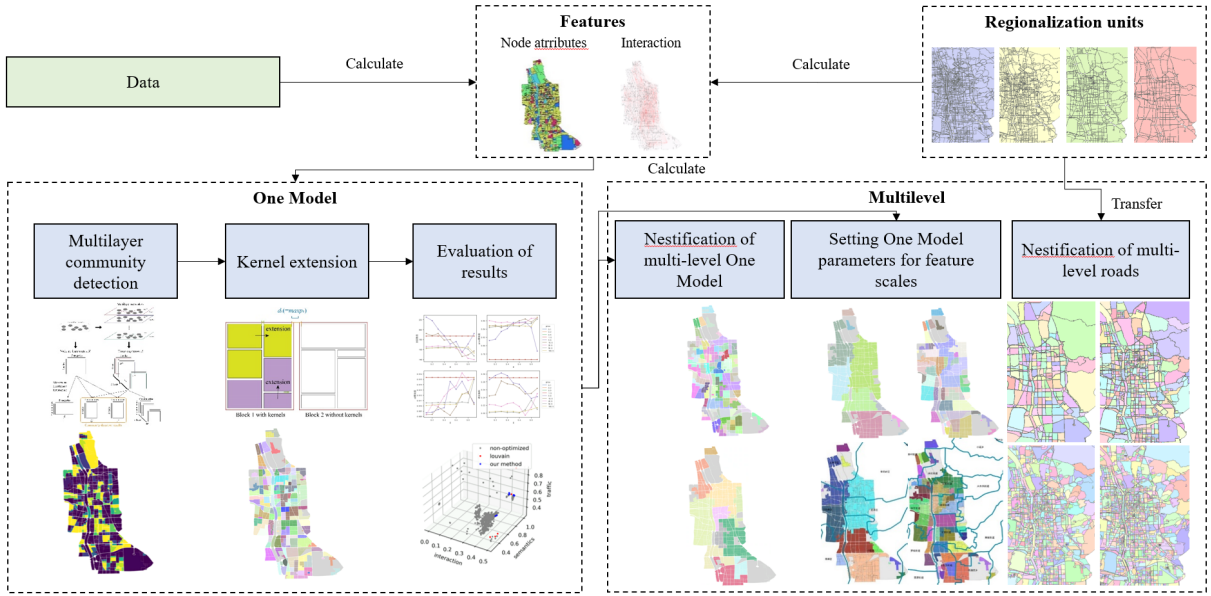

**Fig. 1   Methodological framework**

### 3.1.1 Objective Functions: Attributes vs. Interactions

"Spatial attributes" and "spatial interactions" are two regionalization logics present in the regionalization problem. Spatial attributes refer to the properties inherent to spatial objects. In telecommunications, these include attributes such as the number of users, call volume, and data traffic within the regionalization unit. Regions obtained from "spatial attributes" is referred to as formal regions. This means, based on a certain standard, spatially contiguous areas with similar attributes are aggregated into a region. This aggregation generally requires high similarity within the region and low similarity among different regions [4]. Spatial interaction refers to the interactions that occur between spatial objects, typically characterized by an "intensity" metric. In the realm of telecommunications, spatial interactions can include call volumes between regionalization units or handover volumes caused by users' mobility. Regions derived from "spatial interactions" is termed as "functional regions". This emphasizes that, compared with areas outside the region, there are more interactions among the areas within the region [86], [87].

Given that these two types of regionalization have different optimization directions, the results they produce are also different. Therefore, these two types of data correspond to two different objective functions. This brings challenges to the solution process and highlights the need for a regionalization framework that integrates multiple objectives. Regardless of the types of data, preprocessing must be implemented before regionalization to transform all pre-obtained data to the regionalization units. For instance, for trip origin-destination (OD) data with both starting and ending points grid-based, spatial intersections can be used to allocate the OD of starting and ending grids based on area proportions. This calculates the OD weight between regionalization units, serving as interactive data input for the algorithm.

### 3.1.2 Constraints

#### 3.1.2.1 Size Constraint

Equation (6) indicates that the regionalization units involved in regionalization should be within a certain area range. Semantic and thermal metrics are directly related to area, so oversized or undersized regionalization units can also affect the results and should be excluded and considered as a separate regionalization unit.

#### 3.1.2.2 Morphological constraint

The basic spatial units (shorten as BSU) for regionalization are indicated by constraints (2) ~ (5). They are polygons derived from standardized divisions of a city based on physical quantities like urban morphology, road hierarchy, and natural features [88]. They conform to the city's own spatial organization characteristics, serving to constrain morphology and facilitate operation. By using road networks for graded assignments and spatial processing, different levels of BSU can be obtained. The spacing between two BSU is determined by constrained Delaunay triangulation. Compared to non-overlapping and non-leaking processing methods, this division method effectively avoids some topological issues, captures the adjacency relations between different BSU, and eliminates the impact of data located on roads on the internal structure of the region.

Utilizing BSU and based on the spacing between plots, a depth-first kernel extension search is performed. It not only ensures that the region is as concentrated as possible within a natural geographic area or urban spatial organization range, but also makes a region cluster-structured, thereby avoiding overly irregular regional extensions. This constraint is not placed in the general constraints because it is a constraint generated based on the characteristics of BSU. For other forms of regionalization units, other forms of constraints can be adopted.

## 3.2 Solving Regionalization Schemes

The regionalization issue is modeled as a multi-objective optimization problem. However, due to the vagueness of both the constraints and objective functions, and the solution method being highly influenced by the objective functions and constraints, it is challenging to directly offer a standardized solution method. As previously mentioned, due to the inherent wickedness of spatial problems, it is difficult to provide an absolutely optimal solution in a strict sense. Consequently, we designed a method based on community detection and depth-first search to indirectly obtain a feasible solution [74]. It also provides ports for the input of more kinds of attribute or interaction data in the future and offers adjustable parameters, ensuring algorithmic flexibility to adapt to various different business requirements.

### 3.2.1 Multi-layer Network Community Detection

This community detection step inputs the attribute data of each BSU and the interaction data between the BSU. The output is the membership value of each BSU to each community, ranging from 0 to 1. The sum of the membership values of each BSU to each community is 1.

### 3.2.2 Kernel Extension Based on Depth-First Search

For each BSU, determine its membership degree threshold for every community. BSU with a higher membership value for a particular community are regarded as kernel BSU, while others are viewed as marginal. For marginal BSU, compute the difference in attribute values (such as population heat, traffic heat, and other telecommunication indicators) between them and the kernel BSU within the same block. Identify the marginal BSU that is closest to a core BSU. If this marginal BSU is connected to the kernel BSU, they are merged to form a new kernel BSU. In this way, all BSU within a block are merged into a TAZ, that is, a region. Adjust the parameters according to the characteristic scale to ensure that the regionalization results are comparable to the characteristic scale.

## 3.3 Characteristic Scales and Multilevel Regionalization

There exists some uncertainty in geographical data modeling and analysis. This uncertainty often depends on how observations, spatial processes, and spatial relationships are represented in the analysis [89]. The significance of characteristic scales lies in the inherent spatial scale of influence formed by the physical quantity itself. Different services and their corresponding physical quantities possess different (probably more than one) characteristic scales. The existence of these scales necessitates the introduction of multilevel regionalization to fully incorporate the characteristic scales of different physical quantities, enhancing the aggregation or interaction intensity of physical quantities within the region. For instance, different urban morphological rules have different suitable scales for discussion. When discussing the spatial texture of cities, "plots" are often considered as the smallest meaningful analytical unit [90]. When discussing the relationship between economic activities and urban morphology, it's usually considered at the metropolis level [91]. The theory of spatial containers posits that human activities possess typical characteristic scales, rather than the scale-free distributions commonly revealed in studies. These scales correspond to spatial "containers" that constrain mobility behaviors. Human mobility is organized according to the hierarchical structure of these containers, which correspond to the geographical concept of "places" like communities, cities, urban agglomerations, and regions [14]. Similarly, for the semantics of different regions, the scale of the region is usually finite. A scale that is too large (e.g., the entire city) obviously cannot find a semantic that reasonably summarizes its meaning because the city contains a multitude of different semantics. Focusing on traffic analysis zones, blocks, plots, and buildings allows for the study of their semantics or a mix of several kinds of semantics. Most current researches on identifying semantics of urban functional zones also focus on these scales.

In response to the multi-scale nature of physical quantities themselves, as well as the demand for multilevel management in practical management, we propose three methods for hierarchical division:

1. Directly use the regionalization results of the MNCD-KE algorithm, aggregate each region into a multipolygon as a new regionalization unit, then regionalize it once again with MNCD-KE, aggregating into larger regions to achieve multi-layer nesting.
2. Directly use the nested relationship of BSU. Regard higher-level BSU as higher levels, and the lower-level BSU contained within them as lower levels. The reason for this method is that BSU itself is determined based on city road network levels, and road networks naturally partition urban residents' activities.
3. Obtain regionalization schemes separately based on each characteristic scale. Note that the division results of this method are not fully nested from top to bottom.

Next, we use the Louvain algorithm with a resolution parameter for community detection on the inputted interaction data. By adjusting the resolution parameter, we can detect that with the change of resolution, the scale and number of partitions remain stable within a certain range and undergo abrupt changes at certain points. The trend of these changes is used to determine the characteristic scales.

# 4 Example of TAZ Partitioning and Evaluation of Its Effectiveness

## 4.1 Data Sources

This paper uses the central urban area of Taiyuan, China as a case study for TAZ partitioning. Taiyuan is located in Northern China and is the capital of Shanxi Province. We obtained data for the central urban area of Taiyuan, including roads, administrative boundaries, population, traffic flow, areas of interest (AOI), points of interest (POI), and travel origin-destination (OD). Details of data can be found in Supplementary Material Table S1, and the preprocessing of the data is detailed in Supplementary Material S1.

## 4.2 Objective Functions and Evaluation Indicators

In the telecom domain, information like semantics, population, and traffic are categorized under "spatial attributes", while data like travel OD are part of "spatial interactions". Given the current five-dimensional input, we have categorized them into three groups: semantics, quantity (population and traffic), and interaction (OD and proximity between BSU). We have defined specific objective functions for each category. While both semantics and quantity fall under the "attributes" mentioned in section 3.1.1, they are treated differently in objective calculations since semantics is categorical and quantities are ratio variables. The logic behind defining these objectives is to maximize differences between regions and minimize differences within regions. These objective functions also serve as evaluation metrics for the regionalization solutions.

### 4.2.1 Semantics

For semantics, there is no widely accepted evaluation metric in existing literature. Moreover, once aggregated by region, the larger area makes direct semantic identification challenging. Therefore, we devised our own objective function. The basic idea is the product of intra-region similarity and inter-region dissimilarity. For intra-region similarity, we calculate the entropy of the area of different semantics within the region. A lower value indicates higher similarity within the region. By taking the negative of this value and adding 1, we align the value with the trend of the metric we are measuring. For inter-region dissimilarity, we use an adjacency matrix to calculate the semantic difference between each region and its neighboring regions, weighted by the area of the BSU they contain. The semantics of each region is represented as a vector, constructed from the area ratios of different semantics. After normalizing the area of each semantic type within its group, we use the Euclidean distance to calculate the difference between the semantic vectors of two regions. Given any one region $j$ of the regions in the partitioning schema $Z$, denoted as $Z_C^j$, its area is $A_j$, where the BSU contains a total of $t$ semantic categories. Then the area of the BSU for each semantic category in the region $j$ is respectively $A_j^1, \ldots, A_j^t$. Then its semantic vector normalized by area is $s_j = (\frac{A_j^1}{A_j}, \frac{A_j^2}{A_j}, \ldots, \frac{A_j^t}{A_j})$. A neighboring region recorded by the adjacency matrix of the BSU is $j'$, whose area is $A_{j'}$. The BSU it contains includes a total of $t'$ semantic categories. The area of the BSU for each semantic category is $A_{j'}^1, \ldots, A_{j'}^{t'}$, then its semantic vector normalized by area is $s_{j'} = (\frac{A_{j'}^1}{A_{j'}}, \frac{A_{j'}^2}{A_{j'}}, \ldots, \frac{A_{j'}^{t'}}{A_{j'}})$. Then the objective function is:

$$f_{sem}(Z) = \frac{\sum_{j=1}^{n} A_j * \left( (1 - (-\sum_t \frac{A_j^t}{A_j} \log \frac{A_j^t}{A_j})) \cdot \frac{\sum_{j'} A_{j'} \cdot d(sem_j, sem_{j'})}{\sum_{j'} A_{j'}} \right)}{\sum_{j=1}^{n} A_j} \quad (7)$$

Where $d$ is the inter-vector Euclidean distance. This function represents intra-regional homogeneity through the area-weighted semantic entropy, and inter-regional heterogeneity through the Euclidean distance of semantics between adjacent regions, calculated by area. The function operates similarly to a "convolution" for each region and its surrounding regions based on their areas. A higher value of this function indicates a clearer separation of different semantics brought about by the regionalization.

### 4.2.2 Quantity

For population and telecom traffic, spatial autocorrelation is introduced to measure the variability between regions. Spatial autocorrelation is a way of representing spatial agglomeration relationships in geography. Moran's I index [92] is a statistical measure of global spatial autocorrelation of spatial data with the following formula:

$$I^Z = \frac{n}{\sum_{a=1}^{n}\sum_{b=1}^{n} w_{ab}} \cdot \frac{\sum_{a=1}^{n}\sum_{b=1}^{n} w_{ab}(x_i - \bar{x})(x_j - \bar{x})}{\sum_{a=1}^{n}(x_a - \bar{x})^2} \tag{8}$$

Where $w_{ab}$ is the spatial weights between the regions $a$ and $b$. $x_a$ and $x_b$ are the area-normalized quantities of the regions, and $\bar{x}$ is the average of the area-normalized quantities of all partitions. $n$ is the total number of regions. The value of this value is in the range of [-1,1]. A value greater than 0 indicates a global spatial positive correlation, tending to show features where high and low values cluster together. A value less than 0 indicates a global spatial negative correlation, where high and low values tend to be dispersed. A value of 0 means that there is no global spatial correlation, implying that the distribution of high and low values is random. Based on the purpose of regionalization, we believe that the closer the population and traffic results are to spatial negative correlation, the better the regionalization schema is. Therefore, the objective functions for population and traffic are formulated as follows.

$$f_{pop}(Z) = 1 - I^Z_{pop} \tag{9}$$

$$f_{traffic}(Z) = 1 - I^Z_{traffic} \tag{10}$$

Where $I^Z_{pop}$ and $I^Z_{traffic}$ are the Moran's I index of population and telecom traffic within the regionalization schema $Z$. The trend of the values of the indicators aligns the trend of the regionalization optimization when Moran's I index is subtracted from one.

### 4.2.3 Interaction

For a regionalization scheme $Z$, compute the OD and proximity between each BSU, and construct a network with BSU as nodes and OD or distance as edge weights. The proximity is defined as the spacing among BSU. The community to which each node belongs is determined by the region of the BSU in $Z$. The modularity $Q$ of this network is used as the objective function. The calculation method is:

$$Q^Z = \frac{1}{\sum_r D_r} \sum_{\alpha,\beta} \left[ I_{\alpha\beta} - \frac{D_\alpha D_\beta}{\sum_r D_r} \right] \delta(c^Z_\alpha, c^Z_\beta) \tag{11}$$

$$\delta(c^Z_\alpha, c^Z_\beta) = \begin{cases} 1, & c^Z_\alpha = c^Z_\beta \\ 0, & else \end{cases} \tag{12}$$

Where $I_{\alpha\beta}$ denotes the interaction data (OD or spacing) between the BSU $\alpha$ and $\beta$. $D_r$ denotes the node degree (i.e., the sum of all edge weights connected to the node) computed by edge weights of the BSU $r$. $c^Z_r$ denotes the region to which the BSU $r$ belongs in $Z$.

Then the objective function of the OD ($f_{OD}(Z)$) and proximity among BSU ($f_{prox}(Z)$) are:

$$f_{OD}(Z) = Q_{OD}^Z \qquad (13)$$
$$f_{prox}(Z) = Q_{prox}^Z \qquad (14)$$

Where $Q_{OD}^Z$ and $Q_{prox}^Z$ are modularity when considering OD and proximity among BSU as the interaction in the regionalization schema $Z$, respectively.

## 4.3 Results and Validation

### 4.3.1 Pareto Frontiers of Multidimensional Indicators

Due to the fact that indicators from different dimensions often cannot be optimized simultaneously, some compromises need to be made. We consolidated the five objective functions into three main indicators representing semantics, quantities (including telecom traffic and population intensity), and interaction (travel OD and proximity between BSU). During the consolidation, the average values of the indicators within the same group were calculated. Next, we set a range for each parameter, combined the values of the parameters, and plotted the three-dimensional indicators of the resulting solutions. We extracted the Pareto solutions and grouped them by their number of regions for display. Fig. 2 shows the groups of solutions for the number of regions obtained in Section 4.3.1. Firstly, our algorithm's position in the Pareto space is relatively concentrated, which indicates that the region number ranges based on the Louvain algorithm are also reasonable for our method. Secondly, within each group of region numbers, as the number of regions increases, the number of regions on the Pareto frontier obtained by Louvain algorithm gradually decreases. When the region number exceeds 100, all optimized solutions are derived from our algorithm. Finally, the Pareto solutions given by our algorithm are mostly located in the middle of the Pareto frontier, while those from the Louvain algorithm are mostly at the corner. It shows that our results strike a balance between the three dimensions and are better than the results of the Louvain algorithm that optimizes a single dimension. This advantage is undoubtedly brought about by the multi-dimensional and multi-format input inherent in our algorithm. Methods that can input population, traffic, etc., in a non-discrete manner and simultaneously couple multi-dimensional interactions are very rare in existing regionalization research.

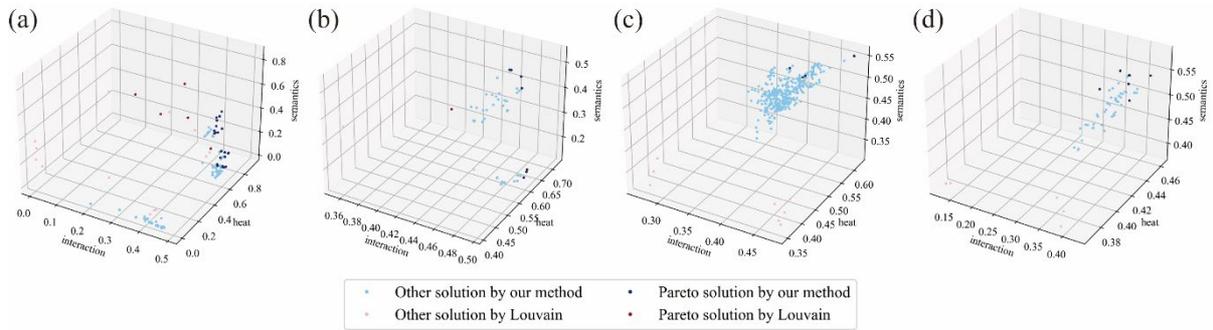

**Fig. 2   Regionalization results under each group of region numbers**

*(a) Less than 50; (b) 50~100; (c) 100~300; (d) Greater than 300. The dark red points denote the Pareto solution given by Louvain algorithm, and the light red points are non-Pareto solution given by Louvain algorithm. The dark blue points denote the Pareto solution given by our algorithm, and the light blue*

*points denote the non-Pareto solution given by our algorithm. The Pareto solution mentioned here is for all regionalization schemas, not for each group of them only.*

### 4.3.2　Characteristic Scales Detected

The spatial distribution of traffic has characteristic scales. It is mainly reflected in the detection of hotspots at two scales, i.e., neighborhoods and streets (Fig. 3). This initially indicates the existence of characteristic scales.

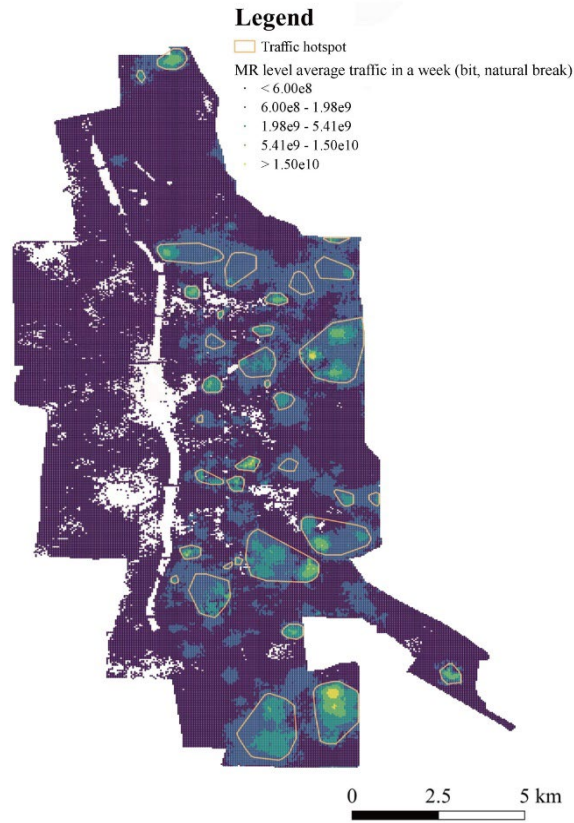

**Fig. 3　Distribution of traffic hotspots in central Taiyuan city**

Further, by using Louvain community detection algorithm and adjusting the resolution parameter $t$, we found that for both the travel OD and BSU distance, the number of communities remains stable within certain resolution ranges. There are evident sudden changes at the boundaries of these ranges (as shown in Fig. 4). The ranges are: less than 50, between 50 and 100, between 100 and 300, and more than 300. Therefore, we can conclude that travel OD and BSU form characteristic scales within these ranges.

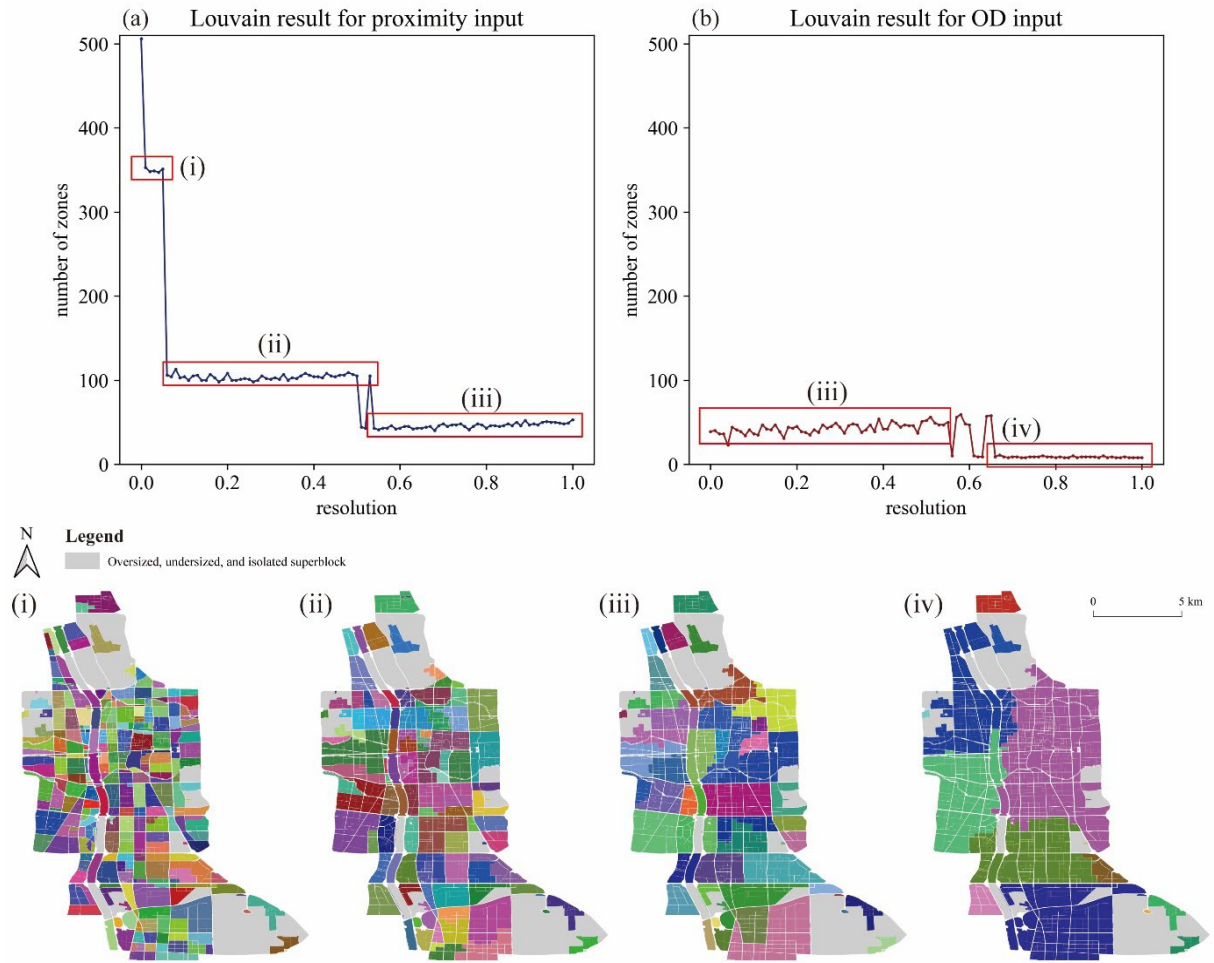

**Fig. 4　The number of regions by adjusting the resolution of Louvain's algorithm**

*(a) traveling OD-based; (b) BSU distance-based. The results for each feature scale are labeled in the red boxes in subfigure (a) and (b), and the numbers correspond to the example feature scale regionalization schemas listed in the subfigures below, including: (i) district level; (ii) sub-district level; (iii) neighborhood level; (iv) community level.*

On the maps, these stable ranges of the number of regions correspond to different spatial scales and constitute the four characteristic scales of regionalization districts (e.g., Fig. 4). We name the four characteristic scales as district level, sub-district level, neighborhood level, and community level, respectively.

### 4.3.3　Discovering High-value Service Regions Based on TAZ

In telecom network planning, we can discover the valuable regions according to the regionalization schema $Z_3$. "Valuable regions" here are defined as certain urban areas that has special functions and need specific kinds of telecom services. Operators can carry out precise services for different valuable regions. For example, Fig. 5 shows some examples of discovering valuable regions.

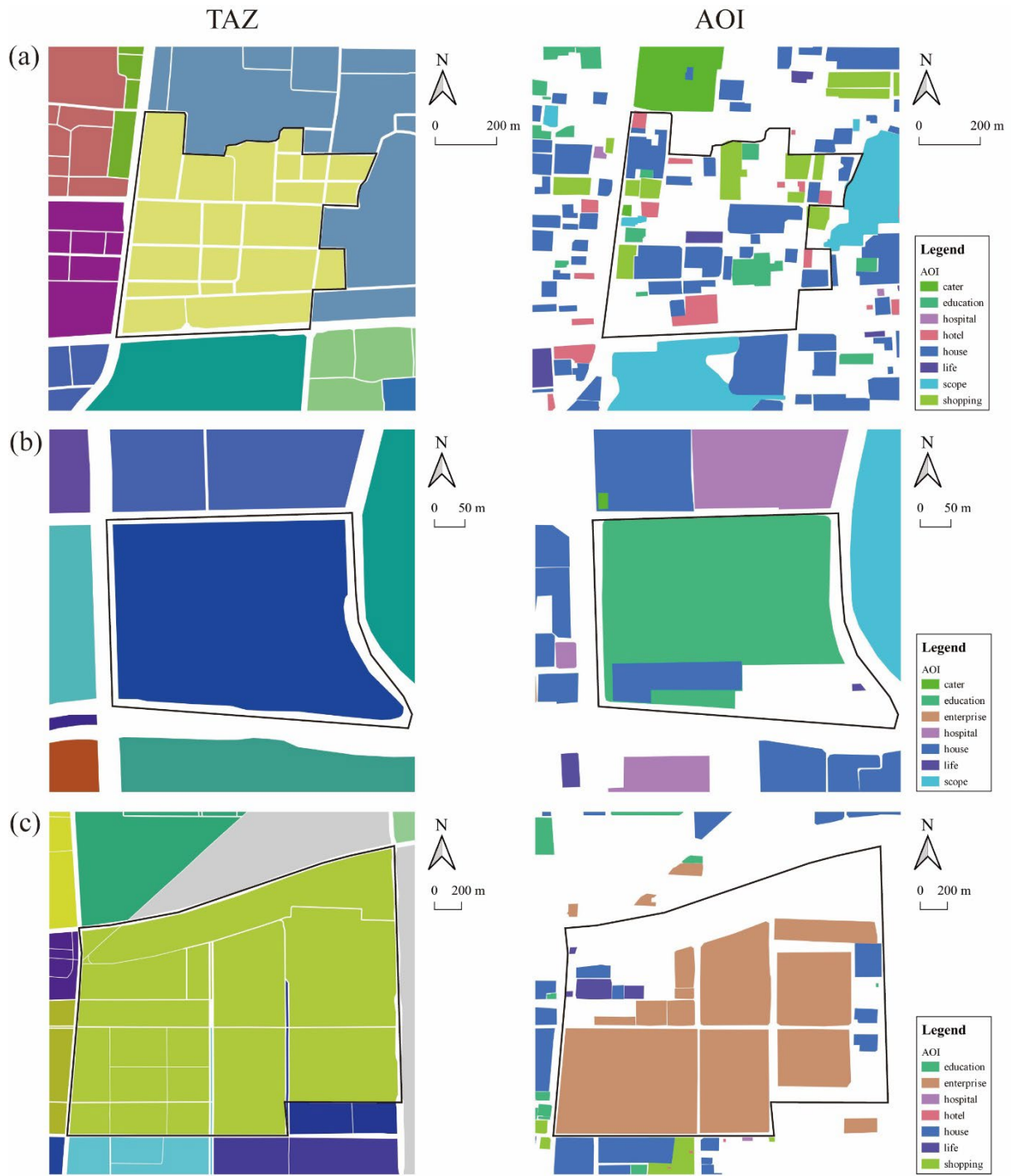

**Fig. 5  Discovering of high-value service regions**

*The ranges of our identified TAZs are marked by black borders. (a)~(c) denote the three identified valuable regions, respectively, with the left column as TAZ and the right column as AOI.*

In Fig. 5, valuable region (a) is located in the center of Taiyuan city and serves as the urban commercial hub, boasting numerous shopping centers and malls. Valuable region (b) is Shanxi Medical University. Valuable region (c) is in the southern part of Taiyuan, home to a large number of factories. These valuable areas require distinct telecommunication services. With accurate identification through TAZ, tailored telecommunication services can be provided to meet the specific needs of each area, such as multimedia and streaming in (a), peak of gaming and entertainment at night in (b), and data transmission of daily production in (c).

### 4.3.4 Comparison for Different Scenarios

We provide comparative options for metric preferences for particular scenarios in regionalization results located at the Pareto frontier. The application scenarios we selected are:

1. User coverage. This implies a preference for regionalization based on population distribution. Therefore, we selected a partitioning scheme with a higher population indicator, corresponding to the scheme $Z_1$ in Fig. 6(a).
2. Crowd mobility coverage. This means as few switching among cells and base stations as possible, which actually requires a schema that emphasizes OD interactions. Therefore, we selected $Z_2$ (Fig. 6(b)), a schema with high interaction metrics.
3. High-value area coverage. This suggests that areas with special semantics should be delineated as much as possible, and therefore regionalization schema $Z_3$ with high semantic metrics was chosen (Fig. 6(c)).

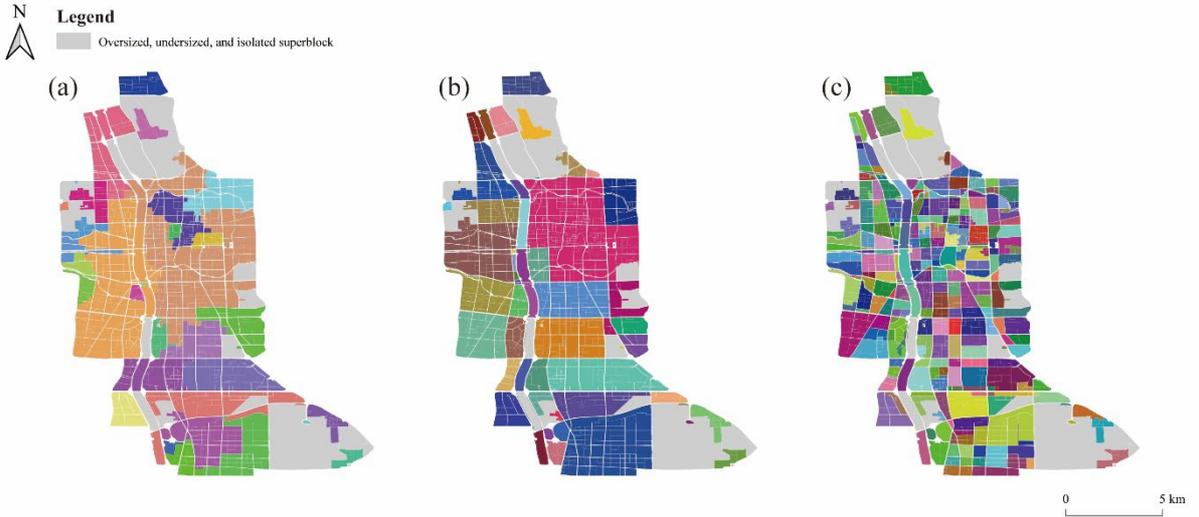

Fig. 6  **Regionalization schemas for each scenario. (a) $Z_1$; (b) $Z_2$; (c) $Z_3$.**

Their numbers of regions and the values of each objective function are shown in Table 2. It can be seen that for their optimization objectives (population for $Z_1$, OD for $Z_2$, and semantics for $Z_3$), the values of their corresponding objective functions are significantly larger than other schemas.

Table 2  **Number of partitions and objective function values**

| $Z$ | Number of regions | $f_{sem}(Z)$ | $f_{pop}(Z)$ | $f_{traffic}(Z)$ | $f_{OD}(Z)$ | $f_{prox}(Z)$ |
| --- | --- | --- | --- | --- | --- | --- |
| $Z_1$ | 36 | 0.126 | 0.776 | 0.765 | 0.080 | 0.810 |
| $Z_2$ | 60 | 0.310 | 0.591 | 0.711 | 0.117 | 0.874 |
| $Z_3$ | 349 | 0.472 | 0.402 | 0.710 | 0.002 | 0.803 |

# 5 Conclusion

Delineating TAZ represents a valuable attempt to apply geographical concepts and methods in the telecommunications industry. Telecommunication activities reflect spatial heterogeneity due to human activities, but they ultimately reveal a unified pattern of human mobility. By delineating TAZ, we effectively "discover" these patterns and delineate them in the form of regionalization. Based on our calculation and case studies, we believe that TAZ can capture the spatiotemporal patterns related to telecommunication activities well. These patterns offer telecom operators a range of regionalization solutions, allowing them to prioritize specific goals. Practically, TAZ facilitates efficient resource allocation and can be applied across various stages, including planning, construction, maintenance, optimization, and marketing, serving as a pillar for the evolution of telecommunications networks towards "autonomous networks". Our examples further indicate that TAZ can play a significant role in identifying valuable regions in telecommunications. Thus, TAZ stands out as a unified and rational regionalization method that meets the needs of the telecommunication industry.